\documentclass[twocolumn,preprintnumbers,amsmath,amssymb]{revtex4-2}
\usepackage{hyperref}
\usepackage{graphicx}
\usepackage{dcolumn}
\usepackage{bm}
\usepackage{comment}
\usepackage{algpseudocode}
\usepackage{physics}
\usepackage{booktabs}

\newcommand{\Er}{E_r}

\begin{document}

\title{Bloch Oscillation Phases investigated by Multi-path St\"uckelberg Atom Interferometry}

\author{Tahiyat Rahman$^1$, Anna Wirth-Singh$^1$, Andrew Ivanov$^2$, Daniel Gochnauer$^1$, Emmett Hough$^1$, and Subhadeep Gupta$^1$}
\affiliation{$^1$Department of Physics, University of Washington, Seattle, WA, USA \\ $^2$Department of Physics, California Institute of Technology, Pasadena, CA, USA}

\date{\today}

\begin{abstract}
Atoms undergoing Bloch oscillations (BOs) in an accelerating optical lattice acquire momentum of two photon recoils per BO. This technique provides a large momentum transfer tool for atom optics, but its full exploitation for atom interferometric sensors requires experimental characterization of associated phases. Each BO involves a Landau-Zener crossing with multiple crossings inducing interference known as St\"uckelberg interference. We develop a multi-path St\"uckelberg interferometer and investigate atomic phase evolution during BOs, up to 100 photon recoil momentum transfer. We compare to numerically calculated single-particle Schr\"odinger evolution, demonstrate highly coherent BO sequences, and assess phase stability requirements for BO-enhanced precision interferometry in fundamental physics and sensing applications.
\end{abstract}

\maketitle

Bloch oscillations (BOs) of cold atoms in an optical lattice \cite{bend96,wilk96} have emerged as a powerful tool in quantum metrology. Local gravity measurements \cite{poli11} and equivalence principle tests \cite{tara14} rely on sensing the external force through measurement of consequent BO frequency, while large momentum transfer (LMT) engineered by optically synthesizing efficient BOs plays a central role in tests of quantum electrodynamics \cite{more20,park18}. In the latter case, LMT-BOs increase the momentum separation of two different atom interferometers (AIs). High-efficiency LMT-BOs {\it within} an AI have the potential to create very large interferometer space-time areas for next generation fundamental physics tests and applications in inertial sensing and gradiometry \cite{tino14}.   

The central appeal for employing BOs for LMT applications compared to other techniques such as pulsed Bragg diffraction \cite{kozu99,gupt01} lies in the high efficiency acceleration possible with BO dynamics restricted to a single lattice energy band, well-separated in energy from other bands. However, this feature is also accompanied by large phase accumulation in this band during the LMT-BO process relative to another band. This can increase demands on experimental controls to maintain phase stability in a BO-enhanced AI composed of paths that lie in these two bands. While 1000$\hbar k$ level momentum transfer has been demonstrated with BOs \cite{more20}, ($\hbar k$ is the lattice photon momentum), BO-enhanced AI has been limited to relatively modest arm-separation ($< 100\hbar k$) \cite{clad09,mull09,mcdo13,page20,gebb21,footkem3}. Continued development of LMT-BO thus relies crucially on the experimental characterization of phases associated with BO processes. Earlier experimental measurements of such differential phases on AI paths have been limited to $N_{\rm BO} \leq 2$ where $N_{\rm BO}$ is the number of BOs \cite{mcal20}. 

Here we investigate the phase associated with LMT-BO processes by utilizing the fact that each BO is accompanied by a Landau-Zener (LZ) crossing which acts as a beamsplitter between the quantum mechanically coupled levels or bands. We use the interference signal from multiple LZ crossings, known as Landau-Zener-St\"uckelberg interference \cite{zene32,stuc32,oliv05,ota18} or simply St\"uckelberg interference, to perform multi-path St\"uckelberg interferometry (MPSI) on a Bose-Einstein condensate (BEC) atom source. The AI paths are composed of atomic wavefunction amplitudes evolving along Bloch bands during an LMT-BO sequence. We find that the MPSI exhibits a characteristic temporal interference pattern from relative St\"uckelberg phase accruals on different AI paths, in very good agreement with coherent, single-particle Schr\"odinger evolution for both ground and excited band BO. Distinct from earlier demonstrations with ultracold atoms that were limited to two-path St\"uckelberg interference \cite{klin10,zene10}, our observations persist even for $N_{\rm BO}=50$, well into the regime of relevance for precision AI.

\begin{figure}
    \center
    \includegraphics[width=0.45\textwidth]{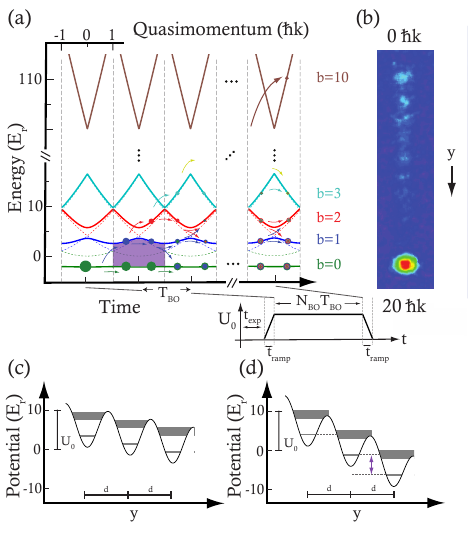}
    \caption{Multi-path St\"uckelberg interferometer scheme. (a) Atomic passage in an extended Brillouin zone picture with avoided band crossings acting as beamsplitters, shown for $U_0/\Er=10$ and $N_{\rm BO}=10$. The shaded region highlights the energy difference between $b=0$ and 1 within one Brillouin zone. Inset shows applied lattice intensity variation. (b) Absorption image ($t_{\rm exp}=3\,$ms, 12ms TOF, 3-shot average) showing MPSI output of 10 momentum states for $N_{\rm BO}=10$, $U_0/\Er=6$, $T_{\rm BO}=0.33(h/\Er)$. Gravity points in the $-y$ direction. (c) Effective spatial potential showing 3 lattice sites for $U_0/\Er=10$ and $T_{\rm BO}=0.5(h/\Er)$. The shading represents the bandwidth. (d) Same as (c) but for $T_{\rm BO}=0.2(h/\Er)$. In this case, the ground state in a site is degenerate with the first excited state in an adjacent site, corresponding to a $2\pi$ phase difference between $b=0$ and 1 during one BO.}
    \label{fig:Fig1}
\end{figure}

Our experimental procedures build on previous work \cite{plot18,goch21} with additional details provided in \cite{SUPP}. Briefly, we produce $^{174}$Yb BECs with $7 \times 10^4$ atoms in an optical dipole trap (ODT). After the BEC is prepared, the ODT is switched off and the atoms are allowed to freely expand for a time $t_{\rm exp}$ before encountering a vertical optical lattice which is adiabatically turned on over a time $t_{\rm ramp}=300\,\mu$s. The value of $t_{\rm exp}$ is chosen to be sufficiently large for all the initial atomic interaction energy to be converted into kinetic energy \cite{cast96}, while keeping the cloud size ($<50\,\mu$m) much smaller than the lattice beams. 

The lattice is formed from a pair of counter-propagating laser beams oriented $5\,$mrad with respect to gravity, with a waist of $1.8\,$mm. The lattice beams are detuned by $\Delta \simeq -3500\Gamma$ from the $556\,$nm intercombination transition, where $\Gamma = 2\pi \times 182\,$kHz, yielding a lattice spacing $d=\pi/k=278\,$nm, and a peak spontaneous scattering rate per $\Er$ of lattice depth of $R_s=2\pi \times 1.1\,$Hz. Here $\Er=h^2/8md^2 = h \times 3.7\,$kHz is the single photon recoil energy. Upon reaching the targeted depth $U_0$, the lattice intensity is kept constant during the interferometry sequence, and subsequently ramped down in $300\,\mu$s. The relative frequency $\delta$ of the lattice laser beams is chirped at the rate $\dot{\delta}_g = 2\pi g/d$ during the lattice intensity ramps to counter the effect of atomic free-fall and provide a stationary lattice in the co-moving frame. Here $g$ is the acceleration due to gravity. During the interferometry sequence, in addition to $\dot{\delta}_g$ we apply a variable frequency chirp $\dot{\delta}_{\rm BO}$ which drives BOs, accelerating atoms relative to the free-fall frame in the $+y$ direction. The adiabatic lattice ramp-down maps the band populations to corresponding populations in free-particle momentum states, which are then observed in time-of-flight (TOF) absorption imaging.

Our MPSI (see Fig.\ref{fig:Fig1}) probes the action of a sequence of $N_{\rm BO}$ Bloch oscillations driven by the frequency chirp $\dot{\delta}_{\rm BO}$ in a lattice of depth $U_0$. In the lattice frame, the atoms experience a periodic potential and an effective constant force which drives BOs. The corresponding tilt of the periodic potential in the freely falling frame is uniquely determined by $\dot{\delta}_{\rm BO}$ and the ensuing BOs have period $T_{\rm BO} = 8\Er/(\hbar \dot{\delta}_{\rm BO})$. The single particle dynamics in this frame are thus described by the Hamiltonian:
\begin{equation}
    \hat{H} = \frac{\hat{p}^2}{2m} + U_0{\rm cos}^2\left(\pi\frac{\hat{y}}{d}\right) - \frac{h}{T_{\rm BO}}\frac{\hat{y}}{d}
    \label{eqn:Eqn1}
\end{equation}
In Fig. \ref{fig:Fig1}(a) we present the MPSI in an extended Brillouin zone scheme. For most of our work, we load the atomic cloud into the ground band $b=0$ of the lattice at quasimomentum $q=0$, before initiating the linear frequency sweep $\dot{\delta}_{\rm BO}$. The initial velocity width is less than $10\%$ of the Brillouin zone. Each avoided crossing acts as a coherent beamsplitter where an interband transition may occur. Equivalently, the atoms experience the Hamiltonian of Eqn.\ref{eqn:Eqn1} during this sweep time set by $N_{\rm BO} \times T_{\rm BO}$. The representative TOF absorption image shown in Fig.\ref{fig:Fig1} (b) displays the different output ports of the MPSI as populations of free-particle states separated by multiples of two photon recoils ($\gg$ velocity width), mapped from the corresponding Bloch band states. Since ground band BOs have one avoided crossing per Brillouin zone, this interference geometry generates $2^{{N_{\rm BO}-1}}$ paths considering only the lowest two bands. Since band numbers up to $b=N_{\rm BO}$ may be populated, this power scaling is a lower bound on the total number of interfering paths. In Figs.\ref{fig:Fig1}(c,d), we show two representative examples of the tilted potential corresponding to two different values of $\dot{\delta}_{\rm BO}$. 

\begin{figure}
    \center
    \includegraphics[width=0.5\textwidth]{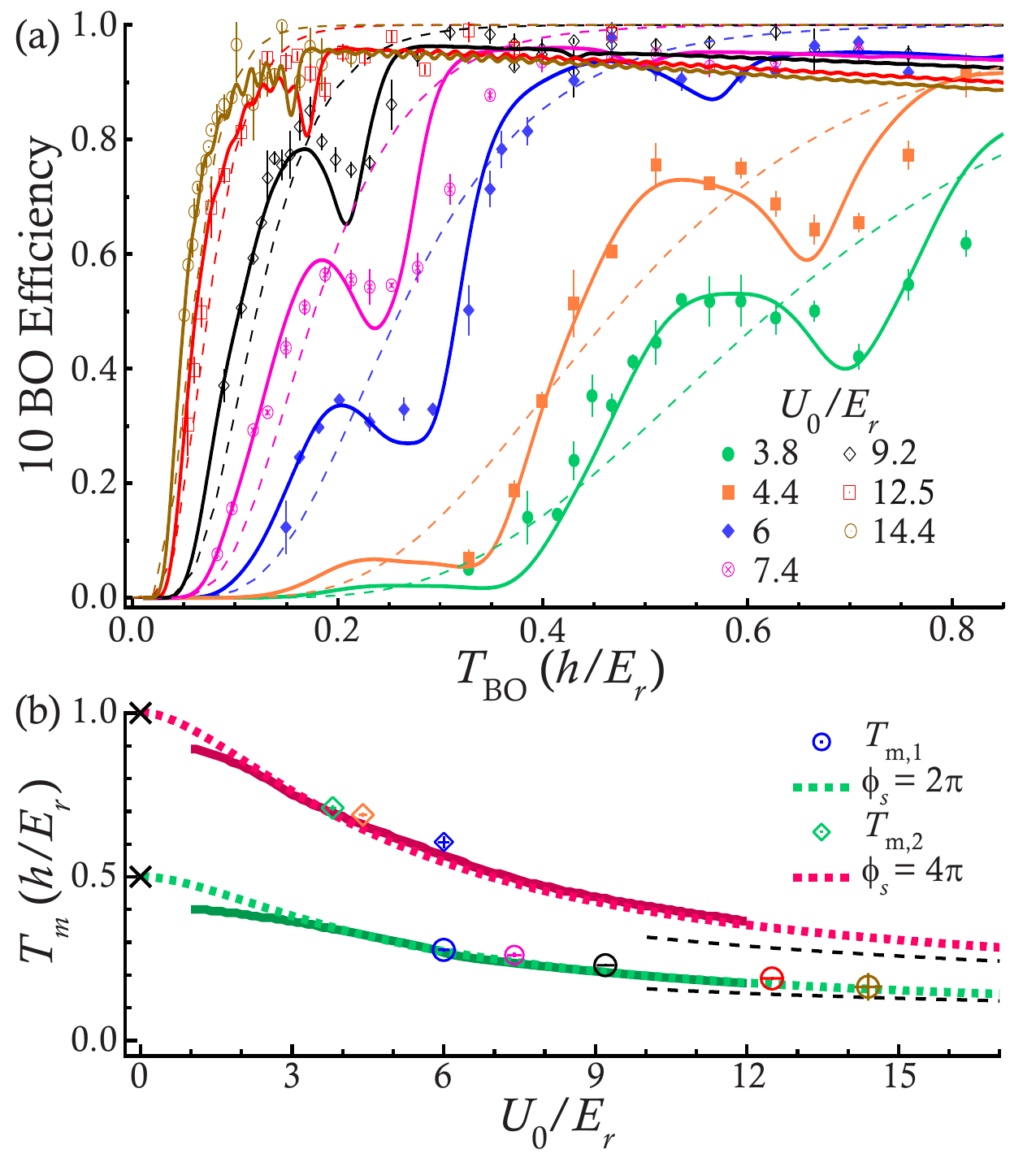}
    \caption{MPSI signal for various lattice depths. (a) Observed 10 BO efficiency variation with $T_{\rm BO}$. Solid lines show the corresponding numerical simulations of Eqn.\ref{eqn:Eqn1}, including spontaneous scattering. Dashed lines show predictions of the Landau-Zener model. (b) Locations of first ($T_{m,1}$) and second ($T_{m,2}$) St\"uckelberg oscillation minima as a function of depth. Markers and solid lines are extracted respectively from experimental data shown in (a) and numerical simulations. Colored dotted lines correspond to setting $\phi_S=2\pi$ and $4\pi$ in Eqn.\ref{eqn:Eqn2} for $b=0$. The $T_{m}$ predictions for a free particle ($U_0 \rightarrow 0$) are shown by the black crosses and for the harmonic approximation ($U_0 \rightarrow \infty$) by the black dashed lines.}
    \label{fig:Fig2}
\end{figure}

We first demonstrate MPSI for ground band BOs with $N_{\rm BO}=10$, a sequence large enough to capture the steady-state $N_{\rm BO} \rightarrow \infty$ behavior. As shown in Fig. \ref{fig:Fig2}(a), the interferometer output, equivalently the 10-BO efficiency, exhibits oscillatory behavior as a function of $T_{\rm BO}$ for various lattice depths. A standard analysis associates each avoided crossing with an LZ tunneling process \cite{zene32,heck02} depleting {\it population} incoherently from $b=0$ leading to the prediction of $[1- {\rm exp}\{-\pi^2E_{\rm bg}^2T_{\rm BO}/(8hE_r)\}]^{N_{\rm BO}}$ for the overall BO efficiency. Here $E_{\rm bg}$ is the depth-dependent band gap \cite{goch19}. As shown by the dashed lines in Fig. \ref{fig:Fig2}(a), the LZ prediction agrees with the observed overall trend with $T_{\rm BO}$ and $U_0$, but fails to capture any of the non-monotonic behavior. 

This non-monotonic behavior is the signature of multi-path St\"uckelberg oscillations. The alternating local extrema at depth-dependent $T_{\rm BO}$ locations correspond to constructive and destructive interference of the contributing paths. To quantitatively understand our observations, we perform numerical simulations of a single particle evolving in the Hamiltonian of Eqn.\ref{eqn:Eqn1} \cite{SUPP}. We additionally incorporate a small contribution from spontaneous scattering with the multiplicative factor ${\rm exp}[-R_s (U_0/2\Er) (\alpha t_{\rm ramp}+N_{\rm BO} T_{\rm BO})]$. Here $\alpha=0.74$ is a numerical factor accounting for the specific shape of intensity ramps used \cite{SUPP}. The good agreement of these simulations (thin solid lines in Fig. \ref{fig:Fig2}(a)) with our observations indicates a high degree of coherence in our MPSI. 

The interference patterns in Fig. \ref{fig:Fig2}(a) represent MPSI measurements of St\"uckelberg phase accrual during BO sequences. For ground band BO, $\phi_S$ is the relative phase between a path which traverses $b=1$ and another which traverses $b=0$ (see Fig.\ref{fig:Fig1}(a)): 
\begin{equation}
\phi^{(b)}_S(U_0)=\frac{\Er T_{\rm BO}}{\hbar} \mathcal{I}^{(b)}
\label{eqn:Eqn2}
\end{equation}
where $\mathcal{I}^{(b)} = \frac{1}{2\Er}\int^{1}_{-1} [E^{(b+1)}(q,U_0)-E^{(b)}(q,U_0)]dq$ and $E^{(b)}(q,U_0)$ is the $q$-dependent energy in band $b$ for lattice depth $U_0$. The location of the pronounced interference minima or depletion in $b=0$ can be estimated by setting the St\"uckelberg phase $\phi_S$ during one BO to an even multiple of $\pi$, corresponding to constructive interference into $b=1$. The first and second minima locations determined by the $T_{\rm BO}$ values that solve $\phi_S = 2\pi$ and $4\pi$ respectively in Eqn. \ref{eqn:Eqn2} are shown as the solid lines in Fig. \ref{fig:Fig2}(b), in clear agreement with the experimentally observed minima locations $T_{m,1}$ and $T_{m,2}$. The dotted lines result from the numerical simulation of Eqn.\ref{eqn:Eqn1} and show excellent agreement with the St\"uckelberg phase calculations using Eqn.\ref{eqn:Eqn2}, except at the lowest depths. The deviations for $U_0 \lesssim 3\Er$ arise from the contribution of the Stokes phase \cite{zene10,shev10,vita99}. 

The fact that Eqn.\ref{eqn:Eqn2} accurately predicts the signal minima emphasizes a resonant effect also found in other multi-path matter-wave interference phenomena such as the temporal Talbot effect \cite{deng99} and the quantum kicked rotor \cite{moor95}. Somewhat surprisingly, these resonances remain pronounced even for large depths $U_0 > 10\Er$ when the avoided crossing locations in momentum space and therefore the beamsplitter timings are less sharply defined due to band flatness. In this regime, intuition can be gained from the position space picture (Fig.\,\ref{fig:Fig1}(c,d)) and spatial tunneling into neighboring sites of the lattice. The position space visualization for $T_{m,1}$ is shown in Fig. \ref{fig:Fig1}(d) and corresponds to resonant tunneling loss into nearest neighbor sites. Likewise $T_{m,2}$ corresponds to resonant tunneling to next-nearest neighbor at half the lattice tilt as $T_{m,1}$.

\begin{figure}
    \centering
    \includegraphics[width=0.5\textwidth]{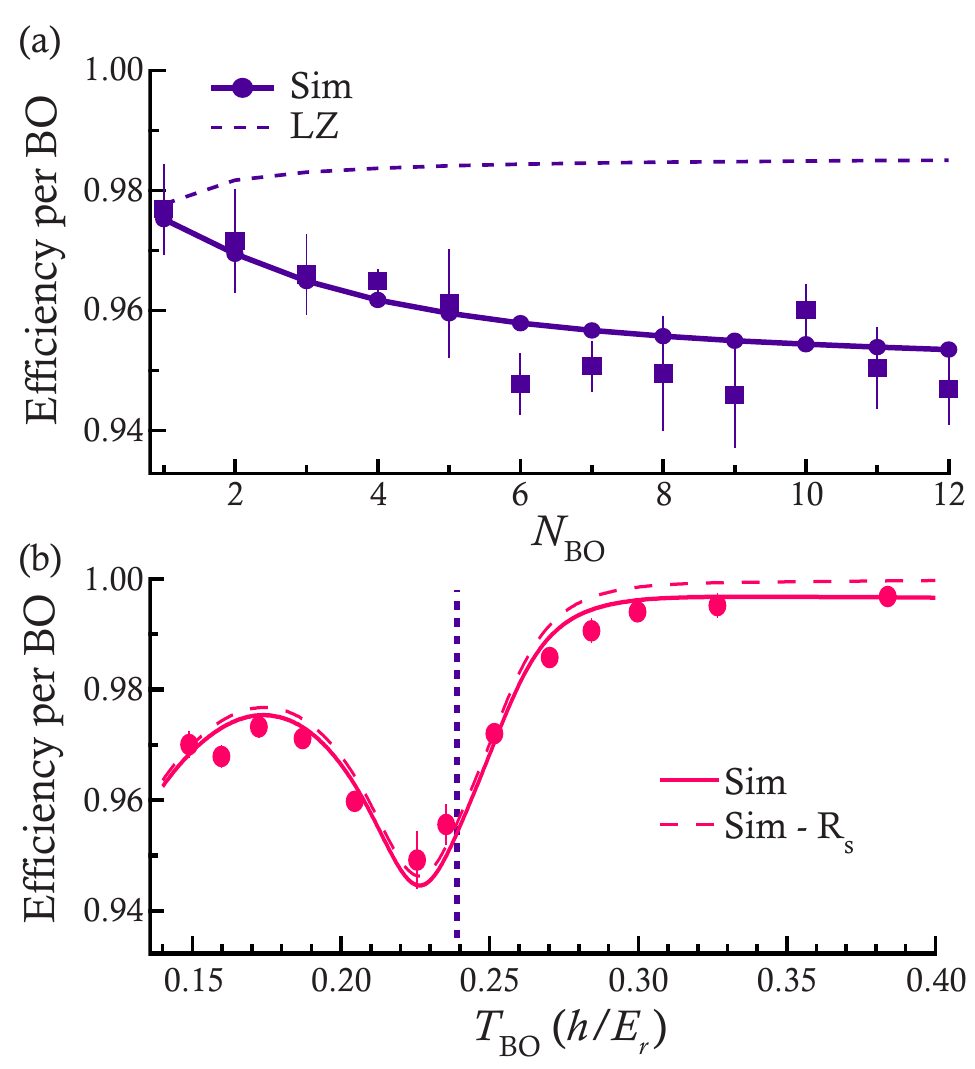}
    \caption{Evolution of St\"uckelberg interference visibility with $N_{\rm BO}$. (a) Per-BO efficiency vs $N_{\rm BO}$ at fixed $T_{\rm BO}=0.24(h/\Er)$ near $T_{m,1}$ for $U_0/\Er=8$. Experimental data are shown as square markers and corresponding numerical simulations including spontaneous scattering as points connected by lines. The dashed line is the Landau-Zener prediction including spontaneous scattering. (b) MPSI signal displayed as per-BO efficiency for an $N_{\rm BO}=50$ LMT-BO sequence with $U_0/\Er=8.3$. The solid (dashed) curve represents the corresponding numerical simulation with (without) spontaneous scattering. The vertical dashed line marks the $T_{\rm BO}$ value in (a).}
    \label{fig:Fig3}
\end{figure}

We now turn to the dependence of the MPSI signal on $N_{\rm BO}$. As schematically shown in Fig.\ref{fig:Fig1}(a) and discussed above, the MPSI involves $ > 2^{N_{\rm BO}-1}$ interfering paths. Even assuming perfectly coherent evolution, this leads to a saturation of the MPSI signal as $N_{\rm BO} \rightarrow \infty$. In Fig. \ref{fig:Fig3}(a), we show the evolution of efficiency per BO for a fixed $T_{\rm BO}$ as $N_{\rm BO}$ is varied. In order to increase the dynamic range for this study, the $T_{\rm BO}$ is chosen near a strong interference feature, leading to an observation of the growth of St\"uckelberg visibility. The signal starts to saturate around $N_{\rm BO}=10$, reflecting the fact that even though the number of paths contributing to it at least double per subsequent BO, the signal in $b=0$ is composed primarily of paths that were generated within the last 10 BOs. The numerical simulation (points joined by lines) is in good agreement with observations. The deviation from the LZ result (dashed line) grows with increasing $N_{\rm BO}$. 

In Fig. \ref{fig:Fig3}(b) we show a representative MPSI signal for $N_{\rm BO}=50$ at a similar depth. Clearly, the visibility remains strong and the agreement with the theoretical model is excellent, with a very small contribution from spontaneous scattering. The vertical dashed line marks the $T_{\rm BO}$ location of Fig. \ref{fig:Fig3}(a). 

\begin{figure}
    \centering
    \includegraphics[width=0.48\textwidth]{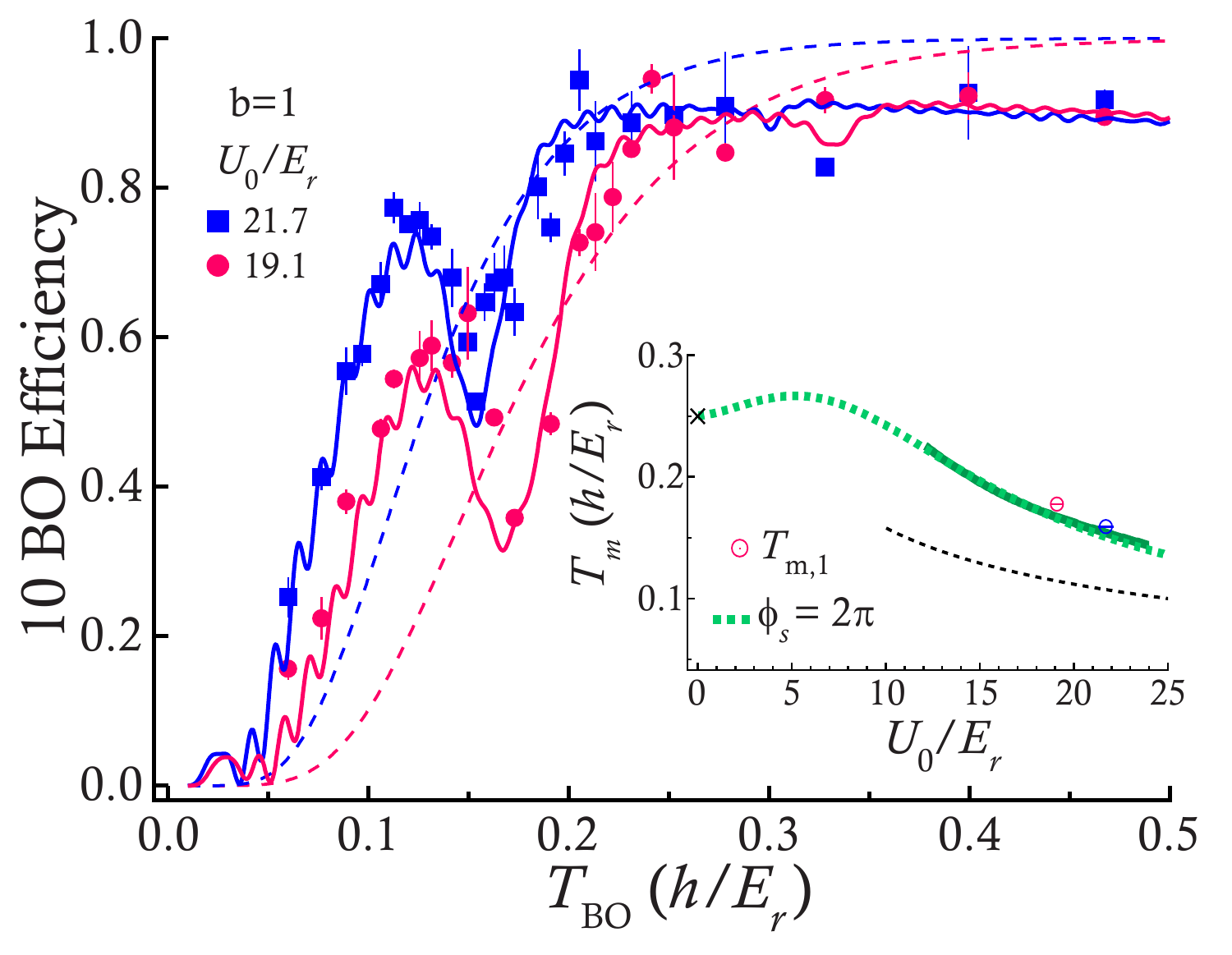}    
    \caption{MPSI signal for excited band. (a) Observed 10 BO efficiency variation with $T_{\rm BO}$ for $b=1$ and two different $U_0$. Solid lines show corresponding numerical simulations of Eqn.\ref{eqn:Eqn1}, including spontaneous scattering. Dashed lines show predictions of the Landau-Zener model. Inset shows location of St\"uckelberg oscillation minima $T_{m,1}$ as a function of depth. Markers and solid lines are extracted respectively from experimental data shown in the main figure and numerical simulations. The colored dotted line corresponds to setting $\phi_S=2\pi$ in Eqn.\ref{eqn:Eqn2} for $b=1$. The $T_{m,1}$ prediction for a free particle ($U_0 \rightarrow 0$) is shown by the black cross and for the harmonic approximation ($U_0 \rightarrow \infty$) by the black dashed line.}
  \label{fig:Fig4}
\end{figure}

Having established the high coherence of long BO sequences using MPSI, we now examine implementation criteria relevant for high precision BO-enhanced AI. For this one needs to consider both the band number separation (=$N_{\rm BO}$) \emph{and} the relative phase evolution $\Phi$ of two paths from an LMT-BO sequence applied on one of the paths. Fast and high efficiency Bloch oscillation will be achieved at high lattice depths. The $U_0/\Er=14.4$ data shown in Fig.\,\ref{fig:Fig2}(a) indicates 0.995 efficiency per $\hbar k$ momentum transfer delivered at 23$\,\mu$s per $\hbar k$ ($T_{\rm BO} \simeq 0.17(h/\Er)$). This is considerably higher than the 0.984 efficiency achieved with optimized Bragg pulses at the same delivery rate of LMT in precision AI with the same atomic species in \cite{plot18}. At $U_0/\Er=25.5$ we measure even higher LMT-BO efficiency of 0.9995 per $\hbar k$ delivered at twice the rate \cite{SUPP}. This corresponds to a 1000$\hbar k$ LMT efficiency of 0.6 delivered in $11\,$ms, with the corresponding Bragg efficiency scaling to the impractical $10^{-7}$. 

Taking 1000$\hbar k$ as a benchmark for next generation high-precision BO-enhanced AIs, we now consider the implications on AI phase from technical considerations of lattice intensity noise. At high depth $U_0 \gg \Er$ and for $N_{\rm BO} \gg 1$, we have $\Phi \simeq \frac{U_0}{2\hbar} T_{\rm BO} N_{\rm BO}$ \cite{SUPP}, and the corresponding phase noise is $\delta \Phi \simeq \epsilon \Phi$ where $\epsilon$ is the relative intensity noise in the lattice beams. For the $1000\hbar k$ example discussed above, achieving a benchmark phase stability $\delta \Phi$ of $100\,$mrad will require $\epsilon = 3 \times 10^{-5}$, which is experimentally challenging. Excited band BOs at ``magic" depths optimized to suppress phase noise induced by lattice intensity variation \cite{mcal20} can help address this challenge.

We finally discuss our experiments on measuring BO phases in excited bands. For this we modify the MPSI sequence by loading the atomic cloud away from avoided crossings at $q=1/2$ in $b=1$ (Fig.\ref{fig:Fig1}(a)) \cite{mcal20}. The results for two different depths are shown in Fig. \ref{fig:Fig4}. The observed St\"uckelberg patterns are similar to those seen in the ground band and are again in good agreement with the numerical simulations. Bandgap considerations suggest that the MPSI signal is dominated by the amplitudes in $b=1$ and 2 rather than $b=1$ and 0, and this is verified by comparing the locations of the interference minima $T_{m,1}$ to the condition $\phi_S=2\pi$ with $b=1$ in Eqn.\ref{eqn:Eqn2} (see Fig. \ref{fig:Fig4} inset). An additional high-frequency oscillation is also seen in the numerical simulation, with a weak amplitude which is below our experimental detection limit. It can be related to the phase accumulation between paths that remain in $b=1$ and $b=2$ throughout the MPSI. 

In summary, we developed a multi-path St\"uckelberg interferometer with ultracold atoms and used it to study the phase imparted during a long sequence of BOs, thus enhancing the toolbox and providing guidance for next-generation fundamental physics tests, gravimetry, and inertial sensing with AI. We demonstrated that BO processes can be applied in a highly coherent manner, implying that the use of large $N_{\rm BO}$ as LMT beamsplitters is limited only by technical effects. The MPSI tool can also help optimize atom-optics parameters in BO-enhanced AIs. Furthermore, we identify phase noise stemming from lattice intensity fluctuations as an important technical challenge for the future, which could be addressed by excited band BOs at magic depths \cite{mcal20}, symmetric AI geometries \cite{page20,gebb21}, or alternative LMT approaches such as Floquet atom optics \cite{wilk22}. 

While completing work on this manuscript, we became aware of related theoretical work in \cite{fitz23}, where spatial Wannier-Stark wavefunctions are used to address the high lattice depth regime relevant for BO-enhanced AI. This approach is in accord with our work and a comparison is included in \cite{SUPP}. 

\begin{acknowledgments}
\textbf{Acknowledgements}: We thank C. Skinner for experimental assistance and A. O. Jamison for a critical reading of the manuscript. We thank F. Fitzek, J.-N. Kirsten-Siemß, N. Gaaloul and K. Hammerer for helpful discussions, and for sharing their Wannier-Stark theoretical calculations. T.R. acknowledges support from the Washington Space Grant Consortium Graduate Fellowship. This work was supported by the National Science Foundation Grant No. PHY-2110164. 

\end{acknowledgments}

\clearpage

\onecolumngrid

\renewcommand{\theequation}{S\arabic{equation}}
\renewcommand{\thefigure}{S\arabic{figure}}
\setcounter{equation}{0}
\setcounter{figure}{0}


\makeatletter
\newcommand*{\addFileDependency}[1]{
\typeout{(#1)}
%
%
\@addtofilelist{#1}
%
\IfFileExists{#1}{}{\typeout{No file #1.}}
}\makeatother

\newcommand*{\myexternaldocument}[1]{%
\externaldocument{#1}%
\addFileDependency{#1.tex}%
\addFileDependency{#1.aux}%
}


\renewcommand{\theequation}{S\arabic{equation}}
\renewcommand{\thefigure}{S\arabic{figure}}

\setcounter{MaxMatrixCols}{10}

\pdfoutput=1

\begin{center}

{\bf \large Supplemental Material for:\\ 
Bloch Oscillation Phases investigated by Multi-Path St\"uckelberg Atom Interferometry}
\end{center}

\twocolumngrid

\section{Experimental Details}

Details of the experimental setup can be found in the main text and in previous work \cite{plot18,goch21}. In this section we provide some additional experimental details supporting the presentation in the main text. \\

\begin{figure}[b]
    \centering
    \includegraphics[width=.45\textwidth]{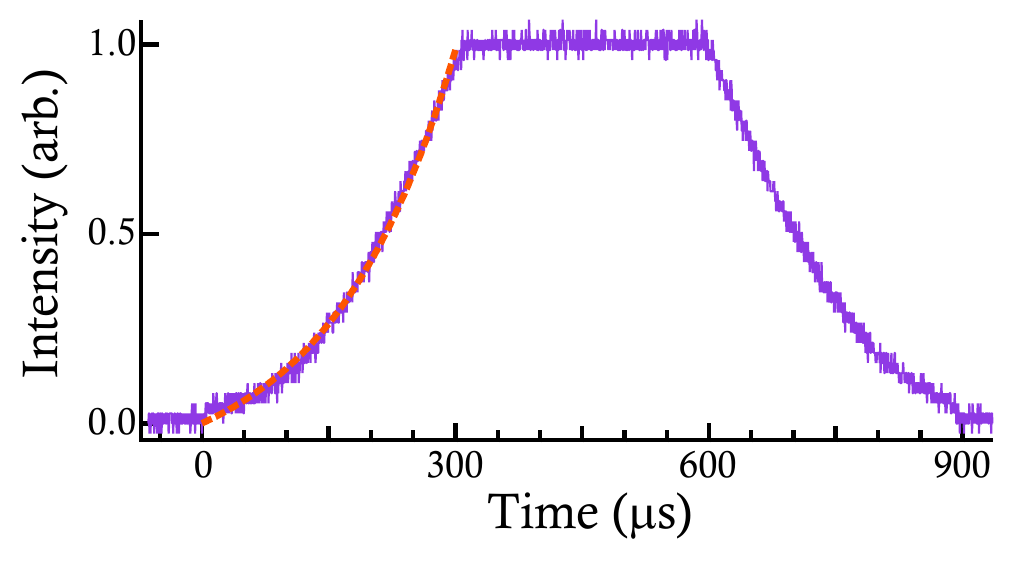}
    \caption{Representative pulse profile for a $T_{\rm BO}=30 \,\mu$s, $N_{\rm BO}=10$, $U_0/E_r=6$ sequence. An exponential fit to the ramp up (dashed line) yields a $1/e$ time of $148$ $\mu$s.}
    \label{fig:FigS1}
\end{figure}

\noindent{\it Optical Lattice}\\
The optical lattice is formed by two counter-propagating laser beams with a detuning of $\Delta/\Gamma \simeq -3500$ relative to the ${^1S}_0 \rightarrow {^3P}_1$ intercombination transition in Yb with wavelength $556\,$nm and linewidth $\Gamma = 2\pi\times182\,$kHz. The relative frequency $\delta$ of the two beams is controlled using acousto-optic modulators that are driven by direct digital synthesizers. The temporal profile of the pulse intensity sketched in Fig 1a of the main text is shown in more detail as a representative oscilloscope trace from a monitoring photodiode in Fig.\,\ref{fig:FigS1}. During the ramp time $t_{\rm ramp}=300\,\mu$s, the lattice depth smoothly rises from 0 to $U_0$, while a gravity-compensating frequency chirp $\dot{\delta}=\dot{\delta}_g$ is applied. The depth is then maintained at $U_0$ for a time $N_{\rm BO}T_{\rm BO}$ while the chirp is set to $\dot{\delta}=\dot{\delta}_g + \dot{\delta}_{\rm BO}$. Finally, the intensity is ramped down to 0 over $t_{\rm ramp}=300\,\mu$s with chirp $\dot{\delta}=\dot{\delta}_g$. Our intensity stabilization keeps the lattice depth constant with $1.9\%$ standard deviation during the experimental sequence. Long-term uncertainty of $\lesssim 10\%$ is observed via calibration using Bragg resonances.

We aligned the lattice to the atoms by optimizing Kaptiza-Dirac diffraction. The orientation of the lattice with respect to gravity was measured to be $5\,$mrad by utilizing the retro-reflection from a cup of water placed on the optical table. 

The chirp $\dot{\delta}_g = 2\pi g/d$ makes the lattice stationary in the co-moving frame of the falling atoms. Gravity points in the $-y$ direction in our experiment (see Fig. 1(b) in the main paper). We determined this chirp experimentally through optimization of Kapitza-Dirac diffraction after long expansion time $t_{\rm exp}$ out of the trap. This yielded $\dot{\delta}_g =-2\pi\times 35.1\,$ kHz/ms, corresponding to $g=9.76$ m/$\rm s^2$ which is $<0.5\%$ below the nominal value of $g$. The corresponding potential systematic on $T_{\rm BO}$ is negligible even for the longest BO periods used in this work. 
\\
\\
\noindent{\it Atomic Interactions}\\
The BEC of $^{174}$Yb formed in the ODT contains about $7 \times 10^4$ atoms, with repulsive mean-field interactions arising from the $s$-wave scattering length $a_s=5.6\,$nm. We measure the associated interaction energy to be $k_B\times 40\,$nK using absorption imaging after a long time-of-flight (TOF) when all of this energy has been converted into kinetic energy \cite{cast96}. The minimum expansion time out of the ODT before the optical lattice is turned on (see Table \ref{tab:Tab1}) is $t_{\rm exp}=3\,$ms, sufficient \cite{cast96} to ensure single particle evolution during the MPSI with negligible interactions.  

\begin{figure}
    \centering
    \includegraphics[width=0.5\textwidth]{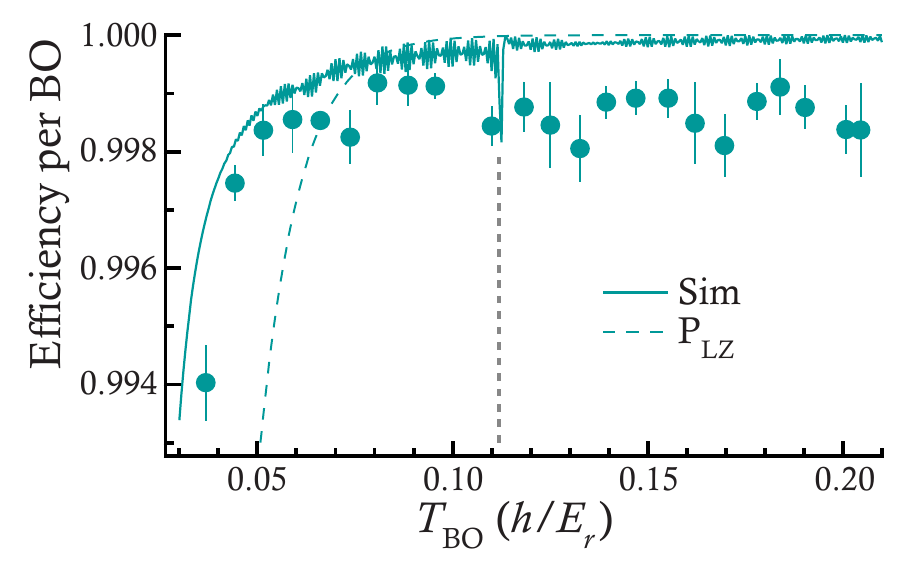}
    \caption{MPSI signal displayed as per-BO efficiency for an $N_{\rm BO}=100$ LMT-BO sequence with $U_0/\Er=25.5$. The solid (dashed) curve shows the corresponding numerical simulation (LZ prediction). The vertical dashed line marks the first St\"uckelberg minimum corresponding to $T_{m,1}$.}
    \label{fig:FigS2}
\end{figure}

The expansion time prior to the lattice turn-on ($t_{\rm exp}$) and the time-of-flight (TOF) after the lattice turn off and before imaging is adjusted depending on $N_{\rm BO}$ due to technical limitations of the imaging setup. In Table \ref{tab:Tab1} we summarize the three different $t_{\rm exp}$ and TOF settings used in this work.
\setlength{\tabcolsep}{0.02\textwidth}
\begin{table}[]
    \centering
    \begin{tabular}{cccc}\toprule
       $N_{\rm BO}$  &   100 BO & 50 BO  & 1-20 BO \\\midrule
          $t_{\rm exp}$ & 9.5 ms& 7 ms& 3 ms \\
          TOF & 1 ms & 4 ms & 12 ms \\\bottomrule
    \end{tabular}
    \caption{Expansion time $t_{\rm exp}$ between ODT switch off and lattice turn on and time-of-flight (TOF) between lattice turn off and absorption imaging used for different sets of experiments reported in this work.}
    \label{tab:Tab1}
\end{table}
\\
\\
\noindent{\it LMT-BO with 100 Bloch Oscillations}\\
In Fig. \ref{fig:FigS2} we show the MPSI signal as per-BO efficiency for a large depth $U_0/E_r=25.5$ and large $N_{\rm BO}=100$, corresponding to a $200 \hbar k$ LMT-BO sequence. St\"uckelberg resonances are present in the corresponding numerical simulation, but with narrow widths. Efficiencies of 0.999 per BO are reached for lattice depth of $U_0/E_r=25.5$ and $T_{\rm BO}=0.08(h/E_r)$ ($2.13\,$ms for 100 BOs). This scales to an overall efficiency of 0.61 for $N_{\rm BO}=500$ ($1000\,\hbar k$ momentum transfer) in $10.7\,$ms.

\section{Phase noise in BO-enhanced AI}
Here we estimate the phase noise induced due to lattice intensity variations in an atom interferometer which uses LMT-BO on one path to achieve high sensitivity in a measurement of interest. In the typical device, the phase of interest will scale with the space-time area enclosed by the AI and therefore to $N_{\rm BO}$, thus the desire for $N_{\rm BO} \gg 1$. However the delivery of large momentum by BOs is accompanied by phase noise from lattice intensity variation during the process. Consider the 2 paths within the AI that accumulate the phase difference of interest due to some external field. For $U_0/\Er \gg 1$ necessary for delivering high efficiency BOs, the ($b=0$) state undergoing ground band BOs is highly localized within lattice sites and experiences the energy shift $\simeq U_0$. The other state is in band  $b\,\gg\,1$ and arrives at $b=N_{\rm BO}$ at the end of the BO process after which the lattice is turned off and relative phase accrual is only from the external field. For sufficiently large $N_{\rm BO}$ (as desired for a high sensitivity application) this other state not undergoing BOs is highly delocalized and experiences the energy shift $\simeq U_0/2$ while the lattice is on. The relative phase accrual during the BOs is then $\Phi \simeq \frac{U_0}{2\hbar}T_{\rm BO}N_{\rm BO}$. Intensity variation of the lattice beams will affect the lattice depth proportionally, leading to a relative depth variation of $\delta U_0/U_0 = \epsilon$. This will lead to AI phase variation $\delta \Phi = \epsilon \Phi$. For the $1000\hbar k$ example discussed earlier, we have $U_0/\Er=25.5$, $T_{\rm BO}=0.08(h/E_r)$, and $N_{\rm BO}=500$. This gives $\Phi \simeq \frac{U_0}{2\hbar}T_{\rm BO}N_{\rm BO} \simeq 3.2 \times 10^3$ as the relative phase from the LMT-BO process. To provide a stability of $\delta \Phi = 100\,$mrad will then require $\epsilon = \delta \Phi/\Phi = 0.1/(3.2 \times 10^3) \simeq 3 \times 10^{-5}$. 

\section{Data Analysis Details}

In this section we provide some data analysis details to support the presentation in the main text.\\

\begin{figure}
    \centering
    \includegraphics[width=0.48\textwidth]{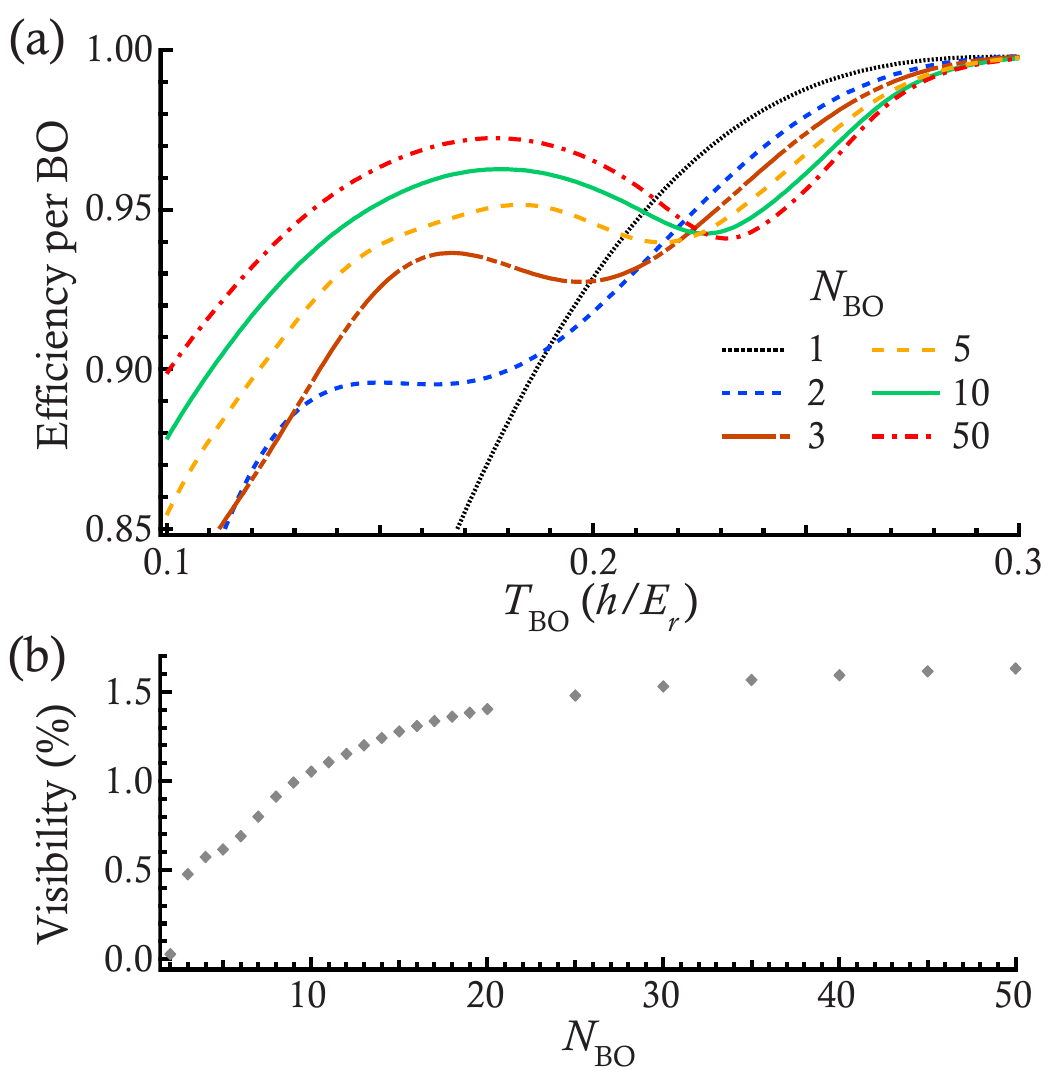}
    \caption{Numerical simulations for different $N_{\rm BO}$ corresponding to the data shown in Fig. 3(a) of the main text for $U_0/\Er=8$. (a) Efficiencies for various $N_{\rm BO}$ up to 50. (b) The visibility of the MPSI signal grows quickly for the first few BOs but the growth slows down for $N_{\rm BO}\gtrsim 10$.}
    \label{fig:FigS3}
\end{figure}

\noindent{\it Numerical Efficiency Curves}\\
Our numerical simulations solve the Hamiltonian shown in Eqn.1 in the main text which describes fully coherent evolution (see details in next Section and Eq.\eqref{Eq:sup_CoherentEvolution}). This yields a BO efficiency or MPSI signal for a given set of BO parameters. We incorporate decoherence from spontaneous scattering of lattice photons with the multiplicative factor ${\rm exp}\left(-R_s(U_0/2\Er)\left(N_{\rm BO}T_{\rm BO}+ \alpha t_{\rm ramp}\right)\right)$ for the overall efficiency. The peak scattering rate at the antinodes is $\frac{(U_0/\hbar)}{(\Delta/\Gamma)} \simeq 2\pi\times1.1 {\rm Hz}\times U_0/E_r = R_s(U_0/E_r)$ calculated in the limit of large detuning $\Delta\,\gg\,\Gamma$. $T_{\rm BO}$ and $t_{\rm ramp}$ are respectively the BO period and ramp on/off times. The factor of 2 assumes that the BEC is spread across several lattice sites experiencing about half of the peak lattice depth. This factor of 2 is the simplest assumption to address the large range of experimental parameters used in this work. Future work can also incorporate the associated spatial wavefunctions. The factor of $\alpha=0.74$ takes into account the smooth intensity ramps that are approximated by exponential shapes (see Fig.\ref{fig:FigS1}). For the efficiency curves shown in the main text (Figs 2(a), 3, and 4), the only parameter adjusted was the peak lattice depth which was optimized by a least squares method while maintaining its value within the $10\%$ long-term variation range observed experimentally. 
\\
\\
\noindent{\it Extraction of St\"uckelberg minima locations}\\
Here we describe how the St\"uckelberg minima shown in Fig. 2(b) and Fig 4 inset of the main text were obtained. For the numerical simulations, we determined the period $T_{\rm BO} = T_{m}$ by generating efficiency curves and locating the corresponding minima via a simple peak finder algorithm. For the experimental data, we fit a region around the minimum using the fit function $1- {\rm exp}(-t/\tau_1) + A[1- {\rm exp}(-(t-t_0)^2/\tau_2^2)]$ which considers a Landau-Zener tunneling shape as background and a Gaussian shape for the resonant dip. This leads to four fit parameters $A$, $\tau_1$, $t_0$ and $\tau_2$, with $t_0$ giving the minimum location. 
\\
\\
\noindent{\it Evolution of MPSI signal with $N_{\rm BO}$}\\
The theoretical curve shown in Fig 3(a) of the main text is obtained from the numerically calculated per-BO efficiency for different values of $N_{\rm BO}$ while keeping $U_0$ fixed at $8\Er$. These numerical simulations are shown in Fig.\,\ref{fig:FigS3}.
As $N_{\rm BO}$ increases, the location of $T_{m,1}$ moves to larger BO periods while the visibility of the St\"uckelberg oscillations grows. We define visibility here as the difference between the efficiencies at the local extrema divided by their sum, in the range of $T_{\rm BO}$ shown. The shift in the minimum location and the visibility both begin to saturate after about 10 BOs. 
\\
\\
\section{Numerical Simulations}

In this section we provide details of the numerical simulations presented in the main text. \\

\noindent{\it Bloch Oscillation Simulation Algorithm}\\ Here we detail the Bloch oscillation simulation algorithm which was used to generate the theory plots in Figs. 2, 3, and 4 of the main paper and Figs.\,\ref{fig:FigS2}, \ref{fig:FigS3}, and \ref{fig:FigS5} of the supplement.

The effect of the quasimomentum sweep can be computed by integrating
the time-dependent Schr\"odinger equation (TDSE), Eq.~(\ref{eq:tdse}), with the time-dependent Hamiltonian
given in Eq.~(\ref{eq:ham}):
\begin{equation}
  \label{eq:tdse}
  i\hbar\partial_t\ket{\Psi} = \hat H(t) \ket{\Psi}
\end{equation}

\begin{equation}
  \label{eq:ham}
  \hat H(t) = E_r\begin{pmatrix}
  (q + 2n)^2 & \nu/4 \\
  \nu/4 & \ddots & \ddots \\
  & \ddots & q^2 & \ddots \\
  & & \ddots & \ddots & \nu/4 \\
  & & & \nu/4 & (q - 2n)^2  \\
  \end{pmatrix},
\end{equation}
where $q=q(t)$ and $U_0 = \nu E_r$ is the lattice depth. The Hilbert space is truncated at $n>N_{\rm BO}/2$ which
reduces its infinite dimensionality to $2n+1$.
   
Let $\hat H(t) = \hat H_0 + \hat D(t)$, where $H_0 = H(0)$. Assuming
$q(0) = 0$,
\begin{equation}
  \hat D(t) = \hat H(t) - \hat H_0 = E_r(q^2\hat I - 4q\hat N),
\end{equation}
where
\begin{equation}
  \hat N = \begin{pmatrix}
    -n \\
    & \ddots \\
    & & 0 \\
    & & & \ddots \\
    & & & & n
   \end{pmatrix}.
\end{equation}

Let
\begin{equation}
  \ket{\Psi(t)} = \sum_i c_i(t)\ket{\psi_i}e^{-i\omega_i t},
\end{equation}
be the state vector in the eigenbasis of $H_0$ with $c_i(t)$ being
some time-dependent complex coefficients. Then, the TDSE
\begin{equation}
  i\hbar\partial_t\ket{\Psi} = (\hat H_0 + \hat D(t)) \ket{\Psi},
\end{equation}
reduces to

\begin{equation}
  \label{eq:csum}
  {\dot c_k} = -\frac{i}{\hbar}\sum_{i,j}[D_j\braket{j}{\psi_i}\braket{\psi_k}{j}e^{-i(\omega_i-\omega_k)t}]c_i.
\end{equation}
where $\ket{j}$ and $D_j$ are the eigenvectors and eigenvalues of
$\hat D$ with the latter running along the diagonal top to bottom.

Eq.~(\ref{eq:csum}) can be written more simply as
\begin{equation}
  {\dot c_k} = -\frac{i}{\hbar}\sum_{i}[\mel{\psi_k}{\hat D}{\psi_i}e^{-i(\omega_i-\omega_k)t}]c_i,
\end{equation}
which leads to the following matrix first-order differential equation
for $C=(c_1, \cdots, c_\mathrm{dim})^T$, where $U=(\psi_1, \cdots,
\psi_\mathrm{dim})$:

\begin{equation}
  \label{eq:odep}
  \begin{aligned}
    \dot C &= -\frac{i}{\hbar} e^{iU^\dag\hat H_0 U t}U^\dag\hat D U e^{-iU^\dag\hat H_0 U t}C \\
    &= -i\Omega_r e^{iU^\dag\hat H_0 U t}U^\dag\hat (q^2\hat I - 4q\hat N)  U e^{-iU^\dag\hat H_0 U t}C \\
    &= -i\Omega_r(q^2\hat I - 4qe^{iU^\dag\hat H_0 U t}U^\dag\hat N  U e^{-iU^\dag\hat H_0 U t})C \\
  \end{aligned}
\end{equation}

\begin{figure}
  \includegraphics[width=0.48\textwidth]{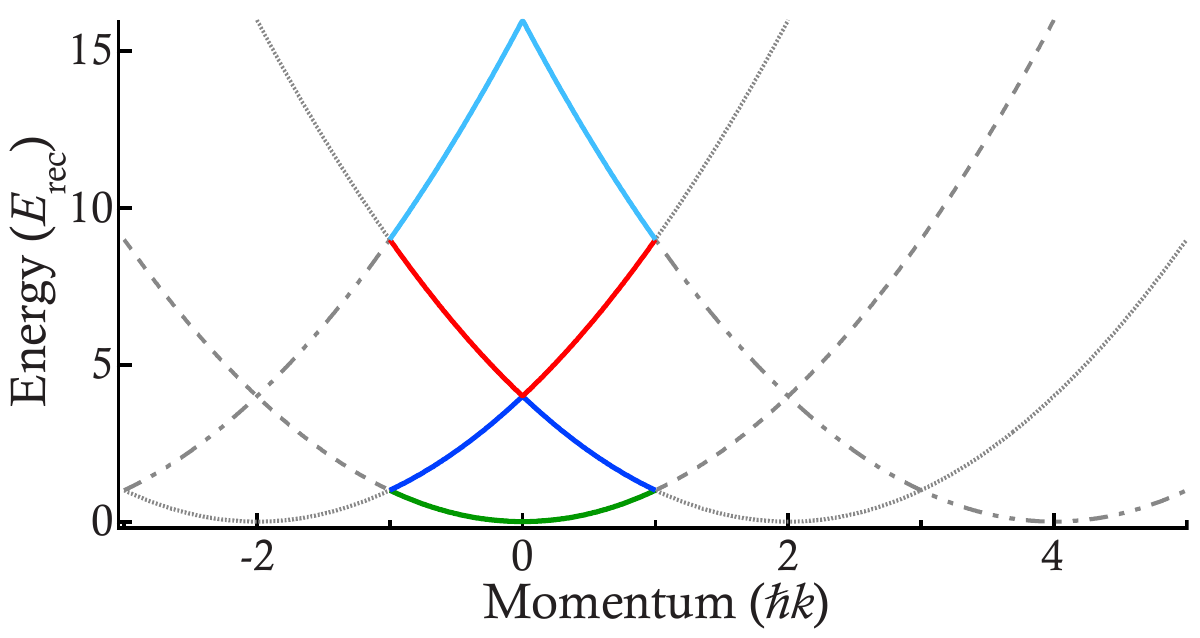}
  \caption{\label{sweep}Extended Brillouin zone picture with parabolas
    corresponding to the dispersion relation of decoupled ($\nu \to 0$) states with solid lines highlighting one Brillouin zone. The
    relationship between diagonal matrix elements of $\hat H(q = 2)$
    and $\hat H(q = 0)$: $H(q = 2)_{i,i} = H(q = 0)_{i-1,i-1}$. This
    means that if $\ket{\psi_i}$ is an eigenstate of $H(q=0)$
    corresponding to $i^\mathrm{th}$ band, then
    $\ket{\downarrow\psi_i}$, where $\downarrow$ denotes a downward
    shift of all state vector components by one position is an
    eigenstate of $H(q=2)$ corresponding to the same band.}
  \label{fig:FigS4}
\end{figure}

We represent one BO as a linear sweep of $q$ from $q=0$ to $q=2$ in
the extended Brillouin zone picture shown in Figure.~\ref{fig:FigS4};
specifically,
\begin{equation}
  q(t) = at = 2t/T_\mathrm{BO}.
\end{equation}

Integrating Eq.~(\ref{eq:odep}) for $t\in[0, T_\mathrm{BO}]$ gives
$c_i$ in the energy basis of $H_0$. To obtain actual band amplitudes
we need to translate them into the energy basis of $H(q = 2)$, which
in practice is very easy: the eigenstates of $H(q=2)$ are
obtained from the eigenstates of $H_0$ with a simple gauge transformation by a downward shift of 1
position for all $\ket{\psi_i}$ (see Figure~\ref{fig:FigS4}).

The unitless equivalent of Eq.~(\ref{eq:odep}) with $t =
(h/E_r)\tau$ is
\begin{equation}
  \label{eq:odeu}
  \begin{aligned}
    \dot C = -i2\pi &q(\tau)\bigg(q(\tau)\hat I \\
    &- 4e^{i2\pi U^\dag
      \tilde H_0 U \tau}U^\dag\hat N U e^{-i2\pi U^\dag\tilde H_0 U
      \tau}\bigg)C,
  \end{aligned}
\end{equation}
where $q(\tau) = 2\tau/\tilde{T}_\mathrm{BO}$ and $\tilde H_0 = \hat
H_0 / E_r$. It can be readily integrated with an adequate ODE solver
such as RK4. Note that the first term in parentheses in
Eq.~\ref{eq:odeu} can be dropped since it only produces a global
phase. Below we summarize the algorithm for simulating $N_\mathrm{BO}$
Bloch oscillations ($c_i^0$ are initial band populations at $t=0$ and
$\mathrm{RK4}$ denotes numerical integration):

\begin{algorithmic}
  \State $C \gets (c_1^0, \cdots, c_\mathrm{dim}^0)$
  \For{$n\gets 1, N_\mathrm{BO}$}
  \State $C(\tilde{T}_\mathrm{BO}) \gets$ \Call{RK4}{$\dot C(\tau)$, $C(0)$, $[0..\tilde{T}_\mathrm{BO}]$}
  \State $\Psi \gets \sum c_i\ket{\psi_i}e^{-i2\pi\tilde\omega_i\tilde{T}_\mathrm{BO}}$
  \ForAll{$\psi_i$}
  \State $c_i \gets \braket{\downarrow\psi_i}{\Psi}$
  \EndFor
  \EndFor
\end{algorithmic}
   
To facilitate an arbitrary initial quasimomentum, Eq.~(\ref{eq:odeu})
is modified to
\begin{equation}
  \begin{aligned}
    \dot C = -i2\pi &(q(\tau) - q_0)\bigg((q(\tau) + q_0)\hat I \\
    &- 4e^{i2\pi U^\dag \tilde H_0 U \tau}U^\dag\hat N  U e^{-i2\pi U^\dag\tilde H_0 U \tau}\bigg)C,
  \end{aligned}
\end{equation}
where $q(\tau) = q_0 + 2\tau/\tilde{T}_\mathrm{BO}$, and the
computational basis $\ket{\psi_i}$ is composed of the
eigenfunctions of $H(q = q_0)$.
\\

\noindent{\it Calculation of the Floquet-Bloch operator} \\
We define the Floquet-Bloch operator \cite{GLUCK2002103} in the basis
of the eigenstates of $\hat H(0)$ as follows:
\begin{equation}
  \hat U_B = \mathcal{T}\left[\exp{-\frac{i}{\hbar}\int_0^{T_\mathrm{BO}}\hat H(t)dt}\right],
\end{equation}
where $\mathcal{T}$ indicates time
ordering of the exponential operation.

Direct evaluation of the Floquet-Bloch operator is quite non-trivial
since, as a consequence of the time-ordering constraint, it requires
computing what's known as Dyson series. However, utilizing the
algorithm described herein, we can very efficiently and with high
accuracy compute $\hat U_B$ indirectly. To that end, let's denote one
iteration of the algorithm with operator $\hat U_1$. Recognizing that
$\hat U_1$ is defined over a vector space of instantaneous Bloch bands
which we denote as $\ket{n}$ (in the preceding discussion,
coefficients $c_i$ represented the amplitudes of the populations of
these bands), we can write
\begin{equation}
  \hat U_B = \sum_{mn}\alpha_{mn}\ketbra{m}{n},
\end{equation}
where $\alpha_{mn}$ are the matrix elements to be determined. Clearly,
\begin{equation}
  \hat U_B\ket{n} = \hat U_1\ket{n} \implies \sum_m{\alpha_{mn}}\ket{m} = \hat U_1\ket{n}.
\end{equation}
Therefore,
\begin{equation}
  \alpha_{mn} = \mel{m}{\hat U_1}{n},
\end{equation}
which means that full evaluation of $\hat U_B$ reduces to $N_{\rm
  max}$ applications $U_1$ operationally defined in the form of the
computational algorithm discussed earlier, with $N_{\rm \max}$ being
the maximal number of the Bloch bands to be accounted for. Once $U_B$
is computed for specific values of lattice depth $\nu$ and Bloch
period $T_\mathrm{BO}$, it can be used for calculating band amplitudes
over an arbitrary number of Bloch periods, $N_\mathrm{BO}$, starting
from an arbitrary initial state $\ket{\Psi_B(t = 0)} =
\sum_nc_n\ket{n}$, as follows:
\begin{equation}
\ket{\Psi_B(t = N_\mathrm{BO}T_\mathrm{BO})} = \hat U_B^{N_\mathrm{BO}}\ket{\Psi_B(t = 0)}.
\label{Eq:sup_CoherentEvolution}
\end{equation}

Of course, the results of this process will be identical to explicit
integration over $N_\mathrm{BO}$ by means of the full computational algorithm;
however, using the Floquet-Bloch operator is considerably (by orders of
magnitude) more efficient and also more accurate than performing
explicit numerical integration over many BO. This gain in efficiency
as well as the ability to study the Floquet-Bloch operator in its
explicit form enable various BO efficiency optimization and analysis
techniques, which will be left to a future study.
\\
\\
\noindent{\it Comparison with Wannier-Stark approach}\\
Here we compare the numerical simulation used in this paper with the Wannier-Stark approach used in \cite{fitz23}. Both sets of numerical simulations are shown in Fig.\,\ref{fig:FigS5} for parameters similar to those in Fig 2(a) of the main paper. We find excellent agreement between the two approaches across a large range of $U_0$ and $T_{\rm BO}$. The deviations between the two appear mostly at low $T_{\rm BO}$ and arise from unavoidable non-adiabaticity in the linear frequency sweep in our model as $T_{\rm BO} \rightarrow 0$, absent in the adiabatic assumption of the Wannier-Stark model of \cite{fitz23}. 

\begin{figure}
    \centering
    \includegraphics[width=0.48\textwidth]{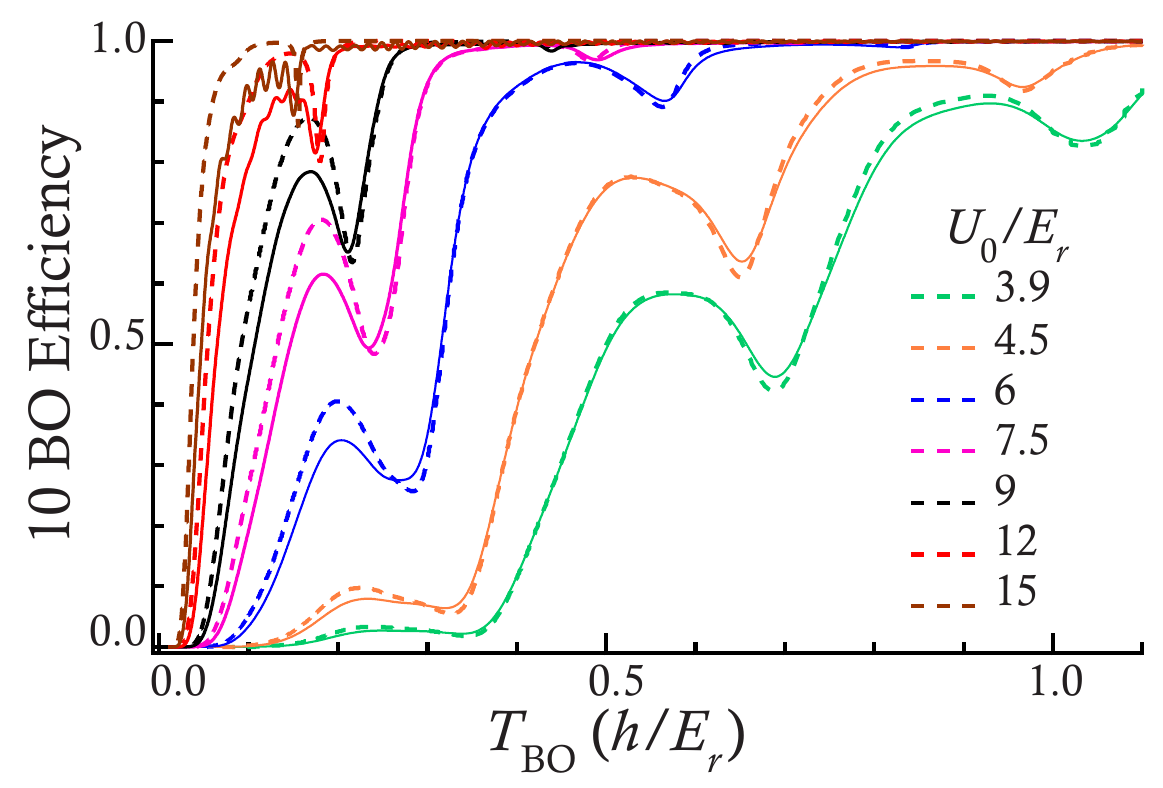}
    \caption{Comparison of the numerical simulations used in this work (solid lines) to the Wannier-Stark approach (dashed lines) of \cite{fitz23} for parameters similar to those in Fig 2(a) in the main paper.}
    \label{fig:FigS5}
\end{figure}


\begin{thebibliography}{34}
\bibitem [{foo()}]{footkem3}%
  \BibitemOpen
  \href@noop {} {\bibinfo  {journal} {Ground-band BOs have been used to
  demonstrate a phase-stable $80\hbar k$ AI in a guided (non free-space)
  geometry \cite{mcdo13}. Recent work in free-space AIs has shown even larger
  momentum splitting with ground-band BOs \cite{page20,gebb21}, however
  phase-stability was demonstrated only up to $20\hbar k$}\ }\BibitemShut
  {NoStop}%
\end{thebibliography}

\end{document}